\documentclass{article}

\usepackage[nonatbib, final]{nips_2017}

\usepackage[utf8]{inputenc} 
\usepackage[T1]{fontenc}    
\usepackage[pdfpagelabels=true]{hyperref}
\hypersetup{
	pdftitle={},
	pdfauthor={AudioLabs},
	pdfsubject={},
	pdfkeywords={},
	pdfview=FitH,
	pdfstartview=FitH,
	bookmarksnumbered,
	colorlinks,
	linkcolor=black,
	urlcolor=blue,
	citecolor=blue
}
\usepackage{url}            
\usepackage{booktabs}       
\usepackage{amsfonts}       
\usepackage{nicefrac}       
\usepackage{microtype}      
\usepackage{amsmath,graphicx,amssymb,bm,units,subfig,acronym,color,enumerate,balance,graphics,graphicx,psfrag,blindtext}
\title{Multi-Speaker Localization Using Convolutional Neural Network Trained with Noise}

\author{
  Soumitro Chakrabarty\\
  International Audio Laboratories\\
  Erlangen\thanks{\,A joint institution of the Friedrich-Alexander-University Erlangen-N\"urnberg (FAU) and Fraunhofer Institute for Integrated Circuits (IIS). E-mail: \texttt{firstname.lastname@audiolabs-erlangen.de}}, Germany \\
  \And
  Emanu\"el A. P. Habets \\
  International Audio Laboratories\\
  Erlangen, Germany \\
}

\begin{document}

\maketitle

\begin{abstract}
The problem of multi-speaker localization is formulated as a multi-class multi-label classification problem, which is solved using a convolutional neural network (CNN) based source localization method. Utilizing the common assumption of disjoint speaker activities, we propose a novel method to train the CNN using synthesized noise signals. The proposed localization method is evaluated for two speakers and compared to a well-known steered response power method. 
\end{abstract}

\section{Introduction}
\label{sec:intro}
In microphone array processing, the source location is an important parameter, which is generally unavailable and needs to be estimated. The location of the source with respect to the array is often given in terms of the direction-of-arrival (DOA) of the sound wave originating from the source position. Over the years, many array processing based methods have been proposed for the task of DOA estimation \cite{Knapp1976, Huang2001c, Brandstein1997, Schmidt1986}. Most of these methods however suffer from degradation in performance in reverberant and noisy conditions \cite{Benesty2008a}.  

Supervised learning methods, being data-driven, provide a distinct advantage for this task, namely they can be adapted to different acoustic conditions via training. If training data from varying acoustic conditions are available, then these methods can also be made robust against adverse acoustic conditions. Recently, several supervised learning methods have been proposed for the task of sound source localization \cite{Takeda2016b, Ma2015, Vesperini2016}.  In \cite{Chakrabarty2017a}, the current authors presented a convolutional neural network (CNN) \cite{Krizhevsky2012, LeCun1998} based supervised learning method for the task of single speaker localization. The CNN was trained with synthesized noise signals, which enabled the creation of large amount of training data in a much more convenient manner than using real-world signals. However, for the case of multi-speaker localization, since the STFT phase components of individual signals are not additive for multiple simultaneously active speakers, the extension of the idea of training the CNN with synthesized noise signals is not straightforward.

To be able to train a CNN for multi-speaker localization using synthesized noise signals, we propose to use the assumption that speakers are not simultaneously active per time-frequency. This assumption is know as W-disjoint orthogonality, has been shown to hold approximately for speech signals \cite{Rickard2002}, and is commonly used in speech separation.

Following a brief introduction to the complete system, we describe the procedure for creating the training data with synthesized noise signals for multi-speaker localization. In addition, we also provide preliminary results from simulated experiments. 
  
\section{Speaker localization with CNNs}
\label{sec:sploc}

Considering an independent source DOA model, we formulate the problem of multi-speaker DOA estimation as a multi-class multi-label classification. The number of classes, $I$, and the class vector is formed based on a discretized set of possible DOA values, similar to \cite{Chakrabarty2017a}.  

The input representation chosen in this work is the same as \cite{Chakrabarty2017a}, where the phase component of the STFT coefficients of the signal are given in the form of a matrix of size $M \times K$, where $M$ and $K$ are the number of microphones and frequency sub-bands, respectively. 
\begin{figure}[t]
	\centering
	\includegraphics[width = 0.97\textwidth]{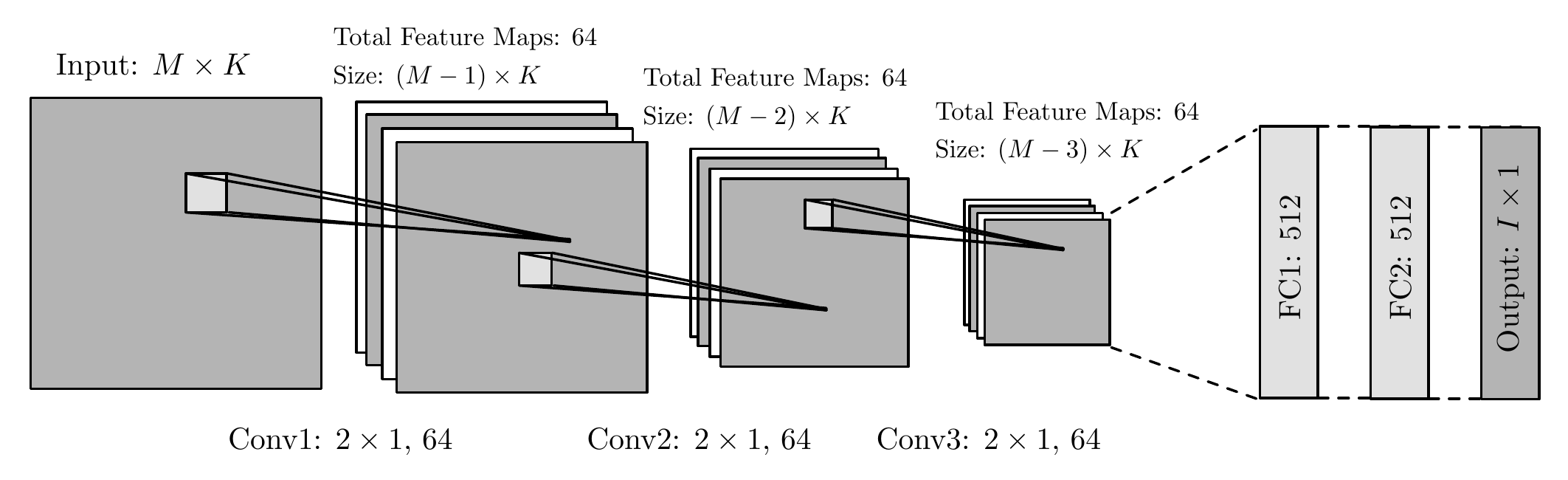}
	\caption{CNN architecture.}
	\label{fig:Arch}
\end{figure}

The convolutional neural network architecture used in this work is shown in Fig.~\ref{fig:Arch}. In contrast to the architecture presented in \cite{Chakrabarty2017a}, we use local filters of size $2 \times 1$, which leads to each filter learning from the phase correlations between the neighboring microphones for each frequency sub-band separately. This is done since we want to utilize the disjoint activity of speech signals for localization. Also, since we formulate our problem as a multi-label classification, the last layer of the CNN now consists of $I$ sigmoid activations. 
  
\section{Generating the training data}
\label{sec:trdata}

For speech signals, it is commonly assumed that the time-frequency (TF) representation of two simultaneously active sources do not overlap. We utilize this assumption to generate training data from synthesized noise signals. In the following, we explain the procedure for generating the training data for two speaker localization per STFT time frame.

As a first step, we generate the training signals for the single speaker case for different acoustic conditions, as explained in \cite{Chakrabarty2017a}. Then, for a specific source array setup, two multi-channel training signals, corresponding to different DOAs, are concatenated along the time axis. Following this, for each frequency sub-band separately, the time-frequency bins for all microphones are randomized to get a single training signal. This procedure is repeated for all combinations of DOAs for all different acoustic conditions considered for training. Finally, the phase map corresponding to each time frame for all training signals is extracted to form the complete training dataset. 

While generating the training data, there are two important things to note regarding the randomization process. First, it is essential that the randomization of the TF bins is done separately for each frequency sub-band, such that the order of the frequency sub-bands remains the same for different time frames. Secondly, it is essential that for each frequency sub-band, the TF bins for all the microphones are randomized together, such that phase relations between the microphones for the individual TF bins are preserved.   

An illustration of this procedure is shown in Fig.~\ref{fig:Pre_res}. From the rightmost figure, we can see that following the randomization procedure, at each time frame there are approximately equal number of TF bins with activity corresponding to the two DOAs. Therefore, at each frequency sub-band of the phase map input to the CNN, the phase of the STFT coefficients for all microphones correspond to a single source. This makes the assumption of disjoint activity of signals implicit within our framework. 

With this training input, the CNN can learn the relevant features for localizing multiple speakers at each time frame from the individual TF bins that contain the phase relations across the microphones for each source DOA separately.  
\begin{figure}[t]
	\includegraphics[width = 0.97\textwidth]{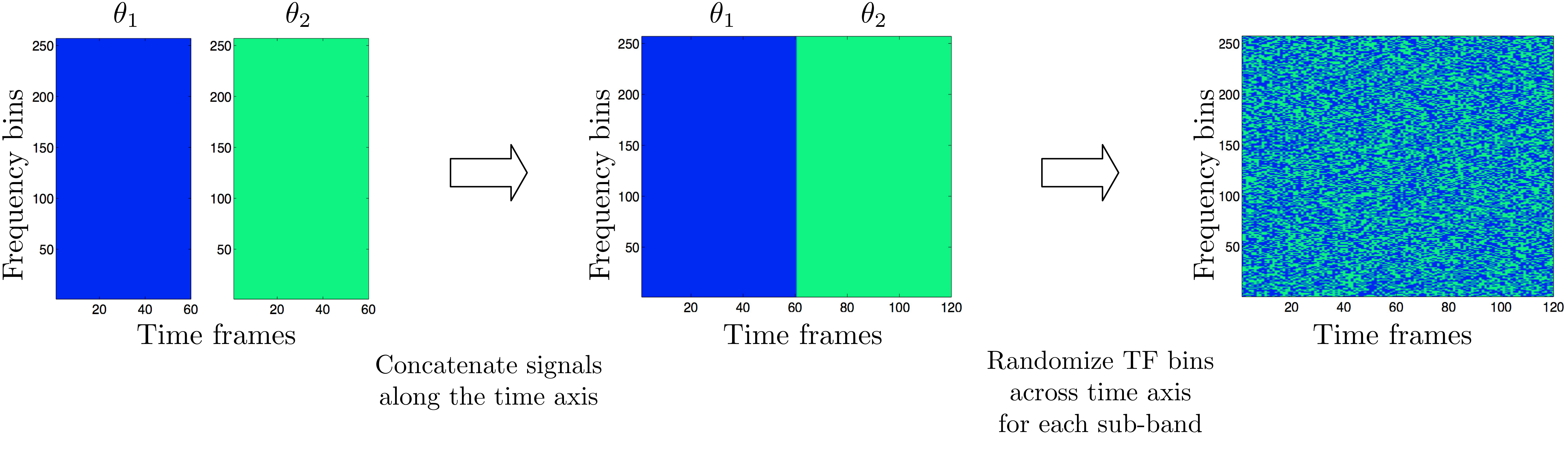}
	\caption{Illustration of the procedure for generating the training data. The STFT representation of two spectrally white noise signals corresponding to DOAs $\theta_1$ and $\theta_2$, of size $N \times K \times M$, where $N$, $K$, and $M$ denote the number of time frames, frequency sub-bands and microphones, respectively are concatenated along the time axis and randomized to get the training signal for the specific DOA combination. From this, for each time frame the phase map is extracted to get the multiple training data samples for this combination. \textbf{Note}: Though the microphone dimension is not shown, the TF bins for all microphones are randomized simultaneously.} 
	\label{fig:Train}
\end{figure}
\section{Experimental results}
\label{sec:exp}
\begin{table}[b]
	\footnotesize
	\centering
	\setlength{\tabcolsep}{1 mm}
	\vspace{1 em}
	\begin{tabular}{c|c}
		\hline \hline
		\multicolumn{2}{c}{{Simulated training data}}	\\
		\hline 
		Signal & Synthesized noise signals \\
		\hline
		Room size  & R1: ($6 \times 6$) m , R2: ($5 \times 4$) m, R3: ($10 \times 6$) m, R4: ($8 \times 3$) m, R5: ($8 \times 5$) m \\
		\hline
		Array positions in room & 7 different positions in each room \\
		\hline
		Source-array distance& 1 m and 2 m  for each position\\
		\hline
		RT$_{60}$ & R1: 0.3 s, R2: 0.2 s, R3: 0.8 s, R4: 0.4 s, R5: 0.6 s \\
		\hline
		SNR & Uniformly sampled from 0 to 20 dB \\		
		\hline \hline
	\end{tabular}
	\vspace{0.5 em}
	\caption{Configuration for training data generation. All rooms are 2.7 m high.}
	\label{tab:Train}\vspace{0.1em}
\end{table}
\begin{table}[t]
	\footnotesize
	\centering
	\setlength{\tabcolsep}{2 mm}
	\vspace{1 em}
	\subfloat[]{
		\begin{tabular}[b]{c|c}
			\hline \hline
			\multicolumn{2}{c}{{Simulated test data}}	\\
			\hline 
			Signal & Speech signals from TIMIT \\
			\hline
			Room size  &  ($9 \times 4 \times 3$) m\\
			\hline
			Array positions in room & 1 arbitrary position\\
			\hline
			Source-array distance& 1.8 m\\		  
			\hline
			RT$_{60}$ &  0.70 s\\
			\hline \hline
		\end{tabular} 
		\label{tab:Test}
	} \hspace{2 em}
	\subfloat[]{
		\vspace{1 em}
		\begin{tabular}[b]{c c | c | c }
			\hline \hline
			SNR & 10 dB & 20 dB & 30 dB \\
			\hline         
			Proposed & $14.3$ & $6.1$ & $1.8$ \\
			SRP-PHAT & $27.1$& $21.6$ & $18.2$  \\	
			\hline \hline
		\end{tabular}
		\label{tab:eval1}\vspace{0.1em}
	}
	\vspace{0.5 em}
	\caption{(a) Configuration for test data. (b) Mean absolute error ($^\circ$) for different levels of spatially white noise.}
\end{table}
We evaluated the performance of the noise trained CNN for the task of DOA estimation of two sources over a complete speech mixture. The posterior probabilities for each DOA class obtained from the CNN output at each time frame are averaged over all the frames, and then normalized to 1. Then the final DOA estimates are obtained by choosing the DOAs corresponding to the classes with the two highest averaged posterior probabilities. The performance was compared to the well-known Steered Response Power with the PHAse Transform (SRP-PHAT) method \cite{Brandstein1997} with similar post-processing applied to the obtained frame-level probabilities. As an objective measure, we used the mean absolute error (MAE) between the true and estimated DOAs over all the speech mixtures in the test dataset.  

We consider a ULA with $M = 4$ microphones with inter-microphone distance of 8 cm, and the input signals are transformed to the STFT domain using a DFT length of 512, with $50\%$ overlap. To form the classes, we discretize the whole DOA range of a ULA with a $5^{\circ}$ resolution to get $I = 37$ DOA classes. The room impulse responses (RIRs) required to simulate different acoustic conditions are generated using the RIR generator \cite{RIRGenerator}. 

The details regarding the training acoustic conditions is given in Table \ref{tab:Train}. The training data is generated as explained in Section \ref{sec:trdata}. Spatially uncorrelated Gaussian noise was added to the training signals with randomly chosen noise levels between 0 and 20 dB before extracting the phase maps. In total, the training data consisted of around 12.4 million time frames for all the different DOA combinations. We used cross-entropy as the loss function and the CNN was trained using the Adam  gradient-based optimizer \cite{Kingma2014}, with mini-batches of 512 time frames. During training, at the end of the three convolution layers and after each fully connected layer, a dropout procedure \cite{Srivastava2014} with a rate of 0.5 was used to avoid overfitting.

To evaluate the performance for all possible DOA combinations, our test dataset consisted of 666 speech signal mixtures, each of length 2 seconds, each corresponding to a specific DOA combination. The acoustic condition for the test case is presented in Table \ref{tab:Test}. 

The results of our preliminary experiment with three different levels of spatially white noise are presented in Table \ref{tab:eval1}. Since the mixture signal for all the DOA combinations was different, we also averaged over the MAE for the two DOAs for each speech mixture. From these results, we can see that the CNN trained with the synthesized noise signals clearly outperforms the SRP-PHAT for all cases. In addition, it is promising to see that even with a simple post processing of the frame-level probabilities, the CNN is able to localize both the sources with such low errors.  

\begin{figure}[tbh]
	\centering
	\includegraphics[width=0.74\textwidth]{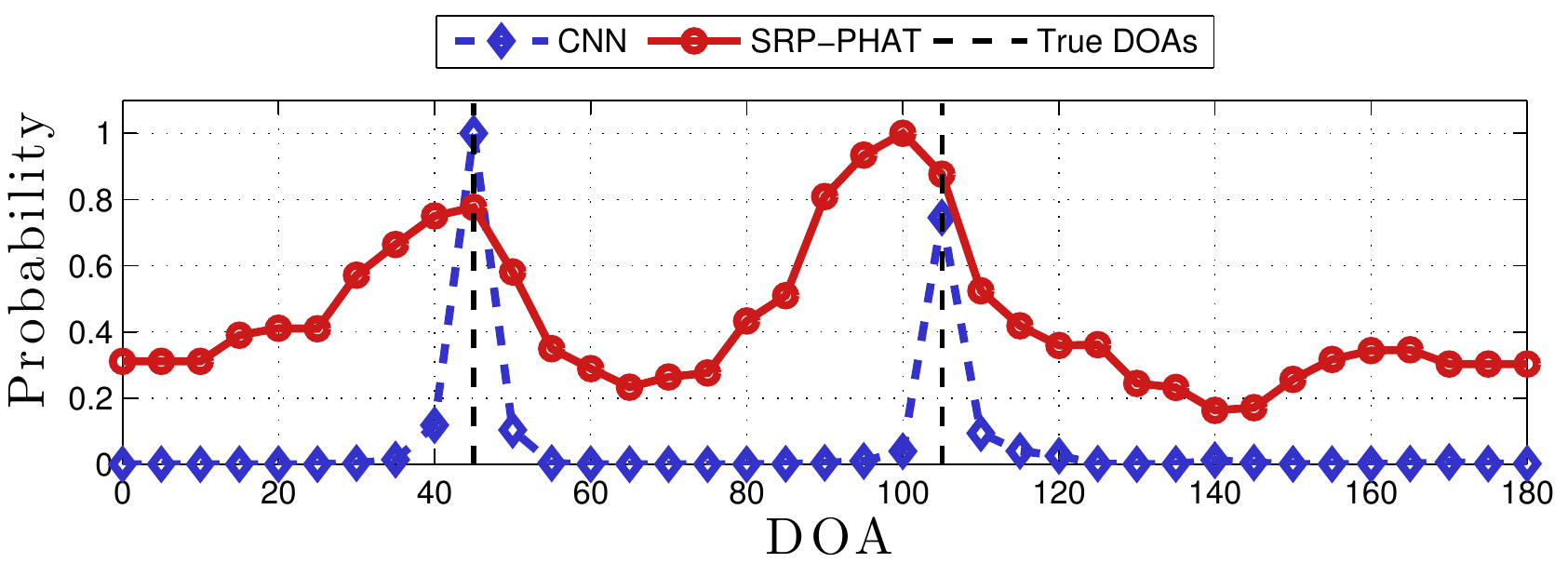}
	\caption{DOA probabilities over the speech mixture when the sources are placed at $45^\circ$ and $105^\circ$.}
	\label{fig:Pre_res}
\end{figure}

In addition, to the MAE, we also show an example result in Fig.\ref{fig:Pre_res}. From the qualitative result, the reason for the big difference in performance between the proposed method and SRP-PHAT becomes clear. It can be seen that for the proposed method there are clear peaks in the distribution whereas for the SRP-PHAT it is much flatter which makes it difficult to obtain accurate final DOA estimates.  

\section{Conclusion and further work}

We presented a CNN-based method trained with synthesized noise signals for the task of multi-source DOA estimation by utilizing the assumption of disjoint activity of speech sources in the STFT domain. Being able to train with synthesized signals enables us to create a large amount of training data conveniently. 

Preliminary experimental results obtained in simulated unmatched acoustic conditions are very promising, especially in terms of the superiority in performance when compared to SRP-PHAT. Future work will focus on more detailed experiments to identify the source of the superior performance as well as limitations of training with synthesized noise signals for this specific task. Finally, we also aim to study the performance of the proposed method in real acoustic conditions.  
\balance
\bibliographystyle{IEEEtran}
\bibliography{sapref_NIPS}

\begin{thebibliography}{10}
\providecommand{\url}[1]{#1}
\def\UrlFont{\rmfamily}
\providecommand{\newblock}{\relax}
\providecommand{\bibinfo}[2]{#2}
\providecommand\BIBentrySTDinterwordspacing{\spaceskip=0pt\relax}
\providecommand\BIBentryALTinterwordstretchfactor{4}
\providecommand\BIBentryALTinterwordspacing{\spaceskip=\fontdimen2\font plus
\BIBentryALTinterwordstretchfactor\fontdimen3\font minus
  \fontdimen4\font\relax}
\providecommand\BIBforeignlanguage[2]{{%
\expandafter\ifx\csname l@#1\endcsname\relax
\typeout{** WARNING: IEEEtran.bst: No hyphenation pattern has been}%
\typeout{** loaded for the language `#1'. Using the pattern for}%
\typeout{** the default language instead.}%
\else
\language=\csname l@#1\endcsname
\fi
#2}}

\bibitem{Knapp1976}
C.~Knapp and G.~Carter, ``{T}he generalized correlation method for estimation
  of time delay,'' \emph{{IEEE} Trans. Acoust., Speech, Signal Process.},
  vol.~24, no.~4, pp. 320--327, Aug. 1976.

\bibitem{Huang2001c}
Y.~Huang, J.~Benesty, G.~W. Elko, and R.~M. Mersereau, ``{R}eal-{T}ime
  {P}assive {S}ource {L}ocalization: {A} {P}ractical {L}inear-{C}orrection
  {L}east-squares {A}pproach,'' \emph{{IEEE} Trans. Speech Audio Process.},
  vol.~9, no.~8, pp. 943--956, Nov. 2001.

\bibitem{Brandstein1997}
M.~S. Brandstein and H.~F. Silverman, ``{A} robust method for speech signal
  time-delay estimation in reverberant rooms,'' in \emph{Proc. {IEEE} Intl.
  Conf. on Acoustics, Speech and Signal Processing (ICASSP)}, vol.~1, Apr.
  1997, pp. 375--378.

\bibitem{Schmidt1986}
R.~O. Schmidt, ``{M}ultiple {E}mitter {L}ocation and {S}ignal {P}arameter
  {E}stimation,'' \emph{{IEEE} Trans. Antennas Propag.}, vol.~34, no.~3, pp.
  276--280, 1986.

\bibitem{Benesty2008a}
J.~Benesty, J.~Chen, and Y.~Huang, \emph{{M}icrophone {A}rray {S}ignal
  {P}rocessing}.\hskip 1em plus 0.5em minus 0.4em\relax Berlin, Germany:
  Springer-Verlag, 2008.

\bibitem{Takeda2016b}
R.~Takeda and K.~Komatani, ``Discriminative multiple sound source localization
  based on deep neural networks using independent location model,'' in
  \emph{IEEE Spoken Language Technology Workshop (SLT)}, 2016, pp. 603--609.

\bibitem{Ma2015}
N.~Ma, G.~Brown, and T.~May, ``Exploiting deep neural networks and head
  movements for binaural localisation of multiple speakers in reverberant
  conditions,'' in \emph{INTERSPEECH 2015}, 2015, pp. 160--164.

\bibitem{Vesperini2016}
F.~Vesperini, P.~Vecchiotti, E.~Principi, S.~Squartini, and F.~Piazza, ``A
  neural network based algorithm for speaker localization in a multi-room
  environment,'' in \emph{IEEE 26th International Workshop on Machine Learning
  for Signal Processing (MLSP)}, 2016, pp. 1--6.

\bibitem{Chakrabarty2017a}
S.~Chakrabarty and E.~A.~P. Habets, ``{B}roadband {DOA} estimation using
  convolutional neural networks trained with noise signals,'' in \emph{Proc.
  {IEEE} Workshop on Applications of Signal Processing to Audio and Acoustics},
  Oct. 2017.

\bibitem{Krizhevsky2012}
A.~Krizhevsky, I.~Sutskever, and G.~E. Hinton, ``Imagenet classification with
  deep convolutional neural networks,'' in \emph{Advances in Neural Information
  Processing Systems 25: 26th Annual Conference on Neural Information
  Processing Systems}, 2012, pp. 1106--1114.

\bibitem{LeCun1998}
\BIBentryALTinterwordspacing
Y.~LeCun and Y.~Bengio, ``The handbook of brain theory and neural networks,''
  M.~A. Arbib, Ed.\hskip 1em plus 0.5em minus 0.4em\relax Cambridge, MA, USA:
  MIT Press, 1998, ch. Convolutional Networks for Images, Speech, and Time
  Series, pp. 255--258. [Online]. Available:
  \url{http://dl.acm.org/citation.cfm?id=303568.303704}
\BIBentrySTDinterwordspacing

\bibitem{Rickard2002}
S.~Rickard and O.~Yilmaz, ``On the approximate {W}-disjoint orthogonality of
  speech,'' in \emph{Proc. {IEEE} Intl. Conf. on Acoustics, Speech and Signal
  Processing (ICASSP)}, May 2002, pp. 529--532.

\bibitem{RIRGenerator}
\BIBentryALTinterwordspacing
E.~A.~P. Habets. (2016) {R}oom {I}mpulse {R}esponse ({RIR}) generator.
  [Online]. Available: \url{https://github.com/ehabets/RIR-Generator}
\BIBentrySTDinterwordspacing

\bibitem{Kingma2014}
D.~P. Kingma and J.~Ba, ``Adam: {A} method for stochastic optimization,''
  \emph{CoRR}, 2014.

\bibitem{Srivastava2014}
N.~Srivastava, G.~Hinton, A.~Krizhevsky, I.~Sutskever, and R.~Salakhutdinov,
  ``Dropout: A simple way to prevent neural networks from overfitting,''
  \emph{Journal of Machine Learning Research}, vol.~15, no.~1, Jan. 2014.

\end{thebibliography}


\end{document}